\documentclass[preprint,aps,superscriptaddress,showpacs]{revtex4-1}
\usepackage{amssymb}
\usepackage{amsmath}
\usepackage{graphicx}
\usepackage{color}

\begin{document}

\title{Current correlations of an on-demand single electron emitter }

\author{A. Mah\'e$^{1}$, F.D. Parmentier $^{1}$, E. Bocquillon $^{1}$, J.-M. Berroir$^{1}$,
D.C. Glattli$^{1, \dag}$, T. Kontos$^{1}$, B. Pla\c cais$^{1}$, G.
F\`eve$^{1, \ddag}$\\
A. Cavanna$^{2}$, Y. Jin $^{2}$ \\
\normalsize{ $^{1}$ Laboratoire Pierre Aigrain, Ecole Normale Sup\'erieure, CNRS (UMR 8551), Universit\'{e} P. et M. Curie, Universit\'{e} D. Diderot}\\
\normalsize{24, rue Lhomond, 75231 Paris Cedex 05,
France,}\\
\normalsize{$^{2}$Laboratoire de photonique et nanostructures, Marcoussis, France.} \\
}

\begin{abstract} In analogy with quantum optics,  short time correlations of the
current fluctuations are measured and used to assess the quality
of the single particle emission of a recently introduced on-demand
electron source. We observe, for the first time in the context of
electronics, the fundamental noise limit associated with the
quantum fluctuations of the emission time of single particles, or
quantum jittering. In optimum operating conditions of the source,
the noise reduces to the quantum jitter limit, which demonstrates
single particle emission. Combined with the coherent manipulations
of single electrons in a quantum conductor, this electron quantum
optics experiment opens the way to explore new problems including
quantum statistics and interactions at the single electron level.
\end{abstract}
\maketitle

Coherent ballistic electronic transport  bears strong analogies
with the propagation of photons.  In particular, the edge states
of a two dimensional electron gas in the quantum Hall regime form
a promising realization of one dimensional ballistic quantum
rails. In this system, electronic interferences have been observed
in Mach-Zehnder interferometers \cite{Ji2003}, using continuous
electron sources based on voltage biased contacts. The electronic
analog of quantum optic experiments \cite{Ol'khovskaya,
Splettstoesser2009}, based on the ultimate control and
manipulation of single electrons in quantum conductors, could be
implemented using the recently proposed single electron emitters
\cite{Feve2007, Blum2007} combined with the development of current
correlation measurements on single electron beams. Furthermore,
these 'electron quantum optic' experiments bear also strong
differences with their photonic counterpart. Electron and photon
statistics differ, and a great richness is also brought by the
presence of Coulomb interaction inducing relaxation
\cite{relaxation} and decoherence
\cite{Roulleau2008} of electronic excitations. In this respect,
single electron emitters offer a new route to study the complex
many-body interaction of a single excitation propagating in the presence of a
Fermi sea \cite{Degio2009}. Some fundamental questions already
arise when one wants to study the elementary processes involved in
the transfer of a single charge from a dot to a one dimensional
lead \cite{Moskalets2008, Keeling2008}. First, the number of transferred charges can
fluctuate (0,1 or 2 ..) if the emitter is not perfect. Another
process, specific to the electronic case, involves the collateral emission of
spurious electron/hole pairs \cite{Keeling2008, Vanevic2008}. It is known from optics that only the short
time intensity-intensity correlations of light $<I(t)I(t+t')>$ can
ensure on-demand emission of a single photon. For perfect single
particle emission, if a particle is detected at time $t$ ($I(t)
\neq 0$), no other particle is detected at time $t+t' \neq t$ and
$<I(t)I(t+t')> \propto \delta(t')$. This so-called Hanbury-Brown
and Twiss (HBT) interferometry  has attracted wide interest in the
characterization of a large variety of single photon emitters
\cite{photon}. In the context of on-demand
electron emitters, these techniques could demonstrate the
realization of a 'clean' emission process resulting in the
emission of a single electronic excitation above the Fermi sea of
the lead at each trigger of the source.

In this paper, we report on the HBT short time correlations
measurements of a periodically driven on-demand electron source
with subnanosecond time control \cite{Feve2007}. In ref
\cite{Feve2007}, the phase resolved measurement of a quantized AC
current in multiples of $2ef_d$, where $f_d$ is the drive frequency,
has brought evidence that the source emits, on average, one
electron followed by one hole at each period of the excitation
signal. Here a breakthrough is reached by the
measurement of the short time autocorrelation (or high frequency
noise) of the current emitted by this electron source. We
demonstrate the existence of two noise limits. The first one
is the standard shot noise associated with the fluctuation of the charge
emitted by the source at each period of the drive. The second one
is a new electronic noise, showing up at high frequency and caused
by the quantum uncertainty in the tunneling escape time of
electrons, which we therefore call quantum jitter. This jitter, or phase noise, is the direct analog of the one observed for triggered single photon
sources \cite{photon}. In
optimum operating conditions of the source,  shot noise disappears
and the current fluctuations reduce to the quantum jitter, demonstrating that
exactly one single particle is emitted at each half period of the
excitation signal. This quantum jitter limit is thus the hallmark of a perfect triggered single particle emitter. Low frequency shot noise suppression has already been observed in pumps. However, the time resolution was not sufficient to reveal the quantum jitter \cite{noisepumps}.
\begin{figure}[!htph]
\centering\includegraphics[scale=0.8]{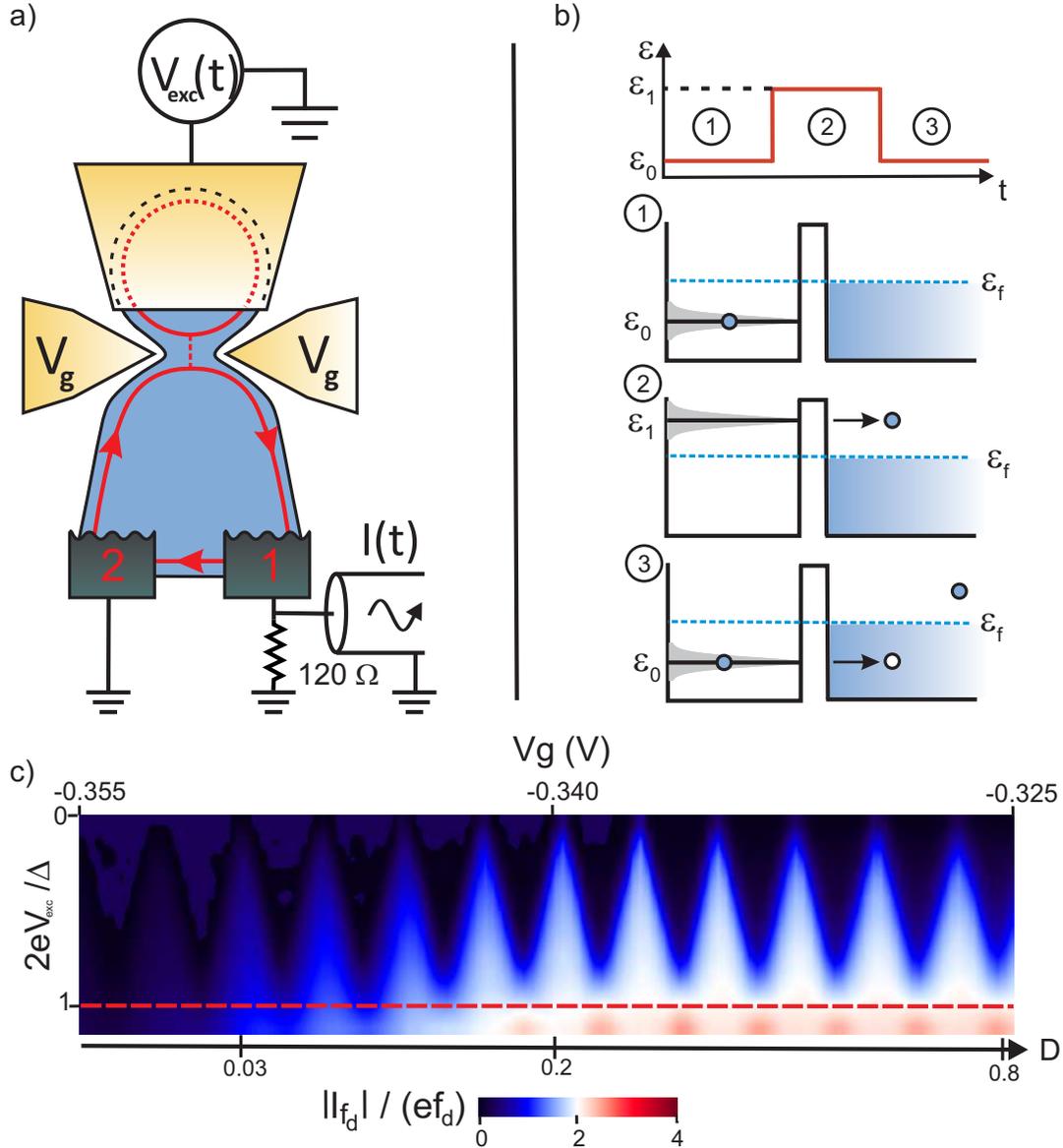} \caption{
 \textbf{a)} Sketch of the
circuit. A single edge state is transmitted between the dot and the
leads with transmission probability $D$ controlled by the QPC gate
voltage $V_g$. Charges emitted by the dot submitted to the
excitation $V_{exc}(t)$ are collected through contact $1$.
\textbf{b)} Sketch of single electron emission as described in the
text.  \textbf{c)} Modulus of the average current first harmonic
$|I_{f_d}|$ in colorscale as a function of the excitation
amplitude and QPC gate voltage.} \label{figure1}
\end{figure}

The source is made of a submicronic quantum dot (see Fig.1.a)
coupled to a two dimensional electron gas (2DEG) by a quantum point
contact (QPC), used as a tunnel barrier of tunable transmission. We
work at a high magnetic field, $B \approx 1.8\,$T, in the quantum
Hall regime with a filling factor $\nu=4$ in the 2DEG leads.  The
QPC gate voltage $V_g$ is set to control the transmission $D$ of the
outermost edge state between the dot and the electron gas, while
inner edge states are reflected. By capacitive coupling, $V_g$ also
controls the static potential of the dot and shifts the position of
the dot discrete spectrum with respect to the Fermi energy. The dot
is also capacitively coupled to a metallic top gate connected to a
high frequency broadband coaxial line. A square ac-voltage
$V_{exc}(t)$ (of peak to peak amplitude $2V_{exc}$) can thus control
the dot potential on subnanosecond timescales with a 20-80 percent
risetime of 60 picoseconds. The dot level spacing $\Delta=4.2\,$K is
responsible for a  finite energy cost for the addition of a single
charge inside the dot (the dot Coulomb energy was found negligible in ref [4], probably
due to the large top gate and is neglected throughout this paper). The
emission of electrons is triggered by the sudden rise ($2eV_{exc}\approx \Delta$) of the dot
potential which brings the last occupied
energy level of the quantum dot (see scheme on Fig.1.b) above the
Fermi energy. It is expected that a single charge is emitted on an
average escape time $\tau=h/ \Delta \times (1/D -1/2) \approx
h/D\Delta$ for $D<<1$ \cite{Feve2007}. By resetting the potential to its initial
value, the dot is reloaded by the absorption of one electron in the
average time $\tau$, leaving a hole emitted in the Fermi sea.
Repeating this sequence at frequency $f_d=1.5 \,$GHz, the periodic
emission (with period $T=1/f_d$) of a single electron followed by a single hole can be
achieved. The charges emitted by the dot are collected in a 120 Ohms
resistor connected to an RF transmission line allowing for the
measurement of the average current and the current noise spectrum
emitted by the dot. As observed in refs \cite{Feve2007,Mahe2008}, the
average current reproduces the exponential time relaxation on a characteristic time $\tau=RC$ of a
classical $RC$ circuit  driven by a square excitation of amplitude $2V_{exc}=C/e$: $\langle I(t) \rangle =\frac{e}{\tau} \frac{e^{-t/\tau}}{1+e^{-T/2\tau}}$ (for $0\leq t \leq T/2$).  For short
escape times $\tau << T/2$, the average emitted charge per half period is quantized: $Q=\int_{0}^{T/2}dt
\langle I(t) \rangle =e$. For escape times comparable to or larger than the half period $\tau \geq
T/2$, electrons do not have enough time to escape which results in a
non-unit emission probability $P<1$ so that $Q=P.e<e$ with $P=\tanh(T/4\tau)$.
In the frequency domain, quantization of the emitted charge shows up in a
quantization of the modulus of the first harmonic of the current
$|I_{f_d}|=2ef_d$ which can be observed on Fig.1.c
representing a color plot of $|I_{f_d}|$ as a function of the
excitation amplitude and QPC gate voltage. White diamonds can be
seen where $|I_{f_d}|=2ef_d$. These diamonds disappear at small
transmission for $\tau \geq T/2$, ($P<1$), and are blurred at large
transmission $D\approx 1$ because of quantum fluctuations of the dot
charge.

\begin{figure}[!htph]
\centering\includegraphics[scale=0.8]{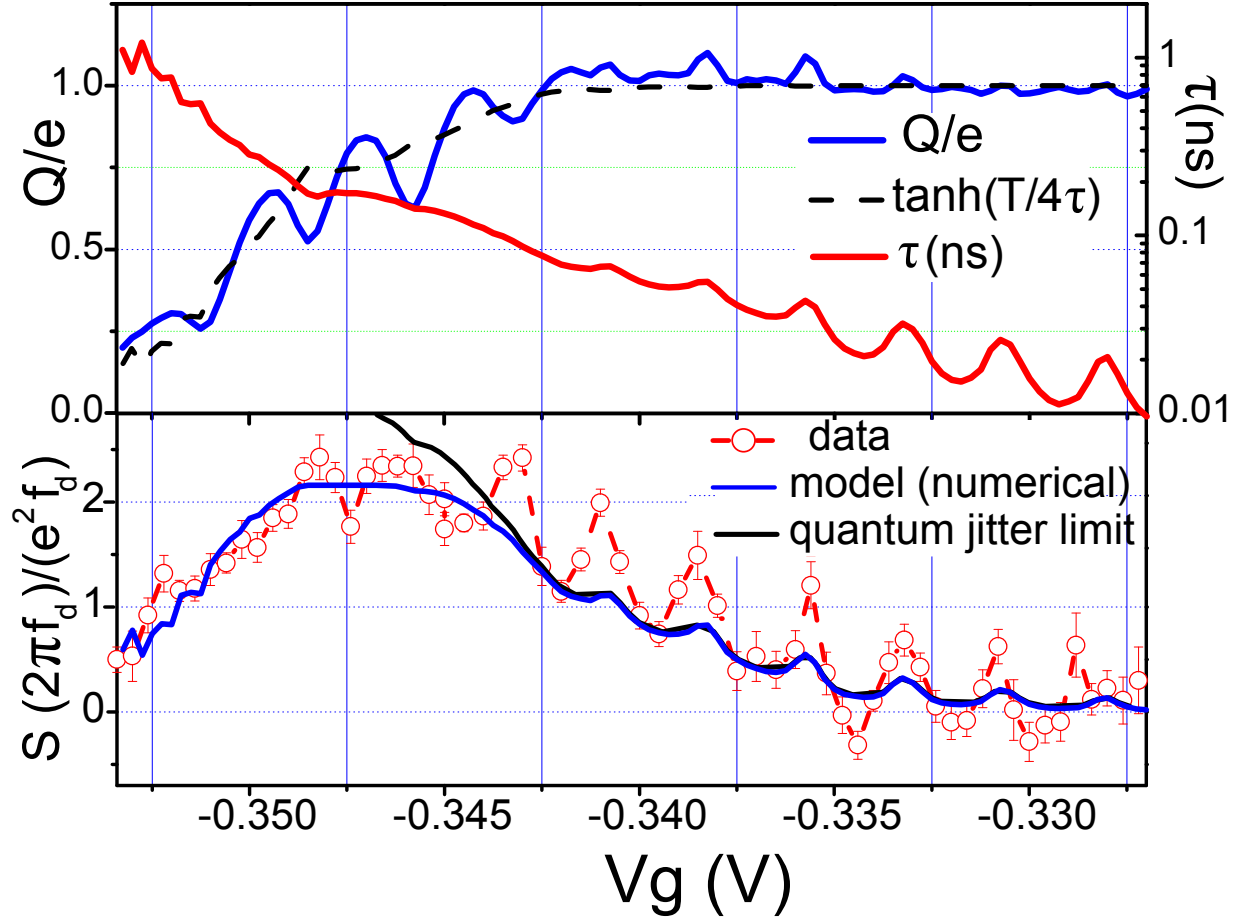} \caption{
\textbf{a)} Average emitted charge $Q$ and escape time $\tau$ for
$2eV_{exc} = \Delta$ as a function of $V_G$. The black dashed line
is the $Q(\tau)$ dependence expected for an exponential relaxation
: $Q=e \times \tanh{\frac{T}{4\tau}}$ \textbf{b)} Experimental
current noise spectrum $S$ as a function of $V_G$ for
$2eV_{exc}=\Delta$ (red points).  The quantum jitter limit $S=4
e^2 f_d \frac{(2\pi f_d \tau)^2}{1 + (2\pi f_d \tau)^2}$, with
$\tau$ given by the experimental data of figure a), has been
plotted in black. The blue trace corresponds to the predictions of
our model.} \label{figure2}
\end{figure}

Although the observation of current quantization is a strong
indication that single charge emission is achieved, the quality of the source can be ascertained by the measurement of the current noise spectrum. In particular, only the latter rule out spurious
multiple particle emission. This is the purpose of this paper
where we focus on high frequency noise measurements for an
excitation amplitude matching the level spacing, $2eV_{exc}=\Delta$
corresponding to the red dashed line on Fig.1.c.

\begin{figure}[!htph]
\centering\includegraphics[scale=0.8]{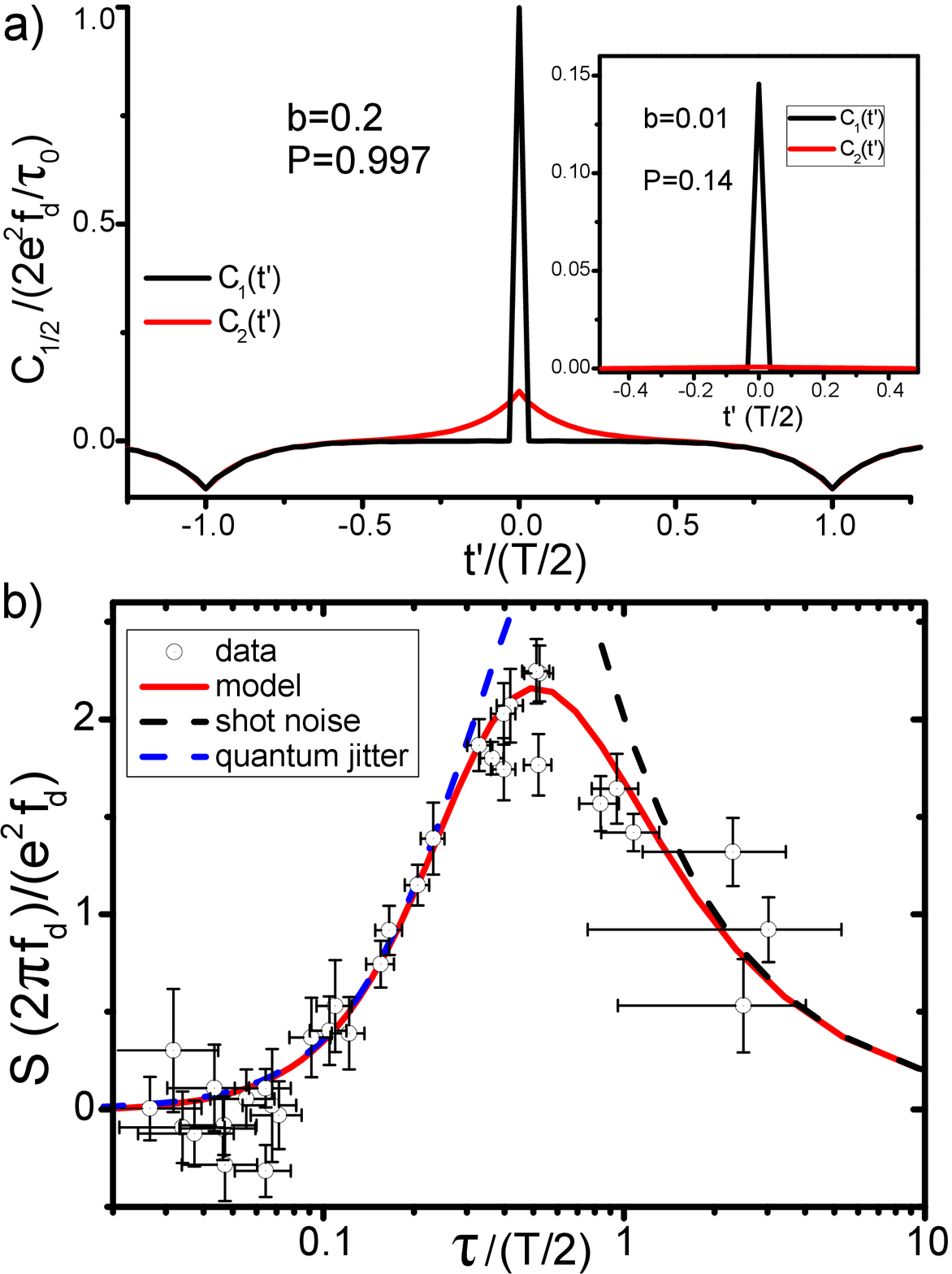} \caption{
\textbf{a)}
 Contributions $C_1(t')$ (black) and $C_2(t')$ (red) to the current correlations $C(t')$ in units of $2e^2f_d/\tau_0$ and as a function of
$t'$ for $P = 0.997$ ($b=0.2$).  Inset:  $C_1(t')$ (black) and $C_2(t')$ (red) for $b=0.01$, $P=0.14$.
 \textbf{b)}
Current noise spectrum $S(2 \pi f_d)$ at the center of the
diamonds plotted as a function of the average escape time $\tau$.
The experimental data are compared to our theoretical predictions
without any adjustable parameter. The asymptotic limits of quantum
jitter and phase noise are plotted in dashed blue and black lines.
}
\end{figure}

Measurements of the high frequency fluctuations of the electron
source differ completely from usual noise measurements in steady
state situations either at low \cite{Kumar96} or high
frequency \cite{zak07}. First, as the circuit is periodically driven,
the statistical average of current fluctuations $C(t,t')=<\delta
I(t) \delta I(t+t')>$ depends as usual on the time difference $t'$ but also
periodically on the absolute time $t$. We will focus in this letter
on the  current correlations averaged, not only on a statistical
ensemble, but also on time $t$, $C(t')= \overline{ <\delta I(t)
\delta I(t+t')>}^t$, and on the noise spectrum $S(\omega) = 2\int dt' C(t')
e^{i\omega t'}$. Second, the intrinsic ac coupling of the circuit
blocks dc current and  the current noise spectrum vanishes at zero
frequency $S(\omega=0)=0$. We have measured the noise power in the
$1.2-1.8\,\mathrm{GHz}$ band centered on the drive frequency $f_d
=1.5 \,\mathrm{GHz}$, excluding the drive frequency using notch
filters. Using an absolute calibration with a thermal source of variable
temperature, we obtain accurate measurements of $S(\omega \approx
 2 \pi f_d)$.

To analyze our experimental results presented in Fig.2., we first
extract $Q$ and $\tau$ from the modulus and phase of the current
first harmonic, considering the exponential dependence of the
average current in time domain. $Q$ and $\tau$ have been plotted
in Fig.2.a as a function of $V_g$. As the average escape time
$\tau$ rises from $20$ $ps$ ($\approx T/30$) to a few $ns$
($\approx 2T$), the emitted charge decreases from a quantized
value $e$ ($P=1$) when $\tau \ll T/2$ to $P.e$ ($P<<1$) when $\tau
\gg T/2$. As can be seen on Fig.2.a., the $P(\tau)$ dependence is very well accounted for by $P=\tanh(T/4\tau)$ (black dashed line).  Having characterized the probability to emit one charge
per half period, we can make a simplified estimation of the
current noise by analogy with low frequency partition noise where
$P$ would be the partitioning probability. In this case, the low
frequency current spectrum $S$ is fully characterized by $P$,
$S=2e \times 2ef_d \times P  (1-P)$ going from shot noise for
$P<<1$ to shot noise suppression for $P=1$. In accordance with our
simple expectation, $S(2\pi f_d)$ presented in Fig.2.b. as a
function of $V_g$ scales in units of $e^2f_d$. It vanishes for
$D\approx 0$, ($P<<1$) and reaches a maximum for $P \approx 1/2$.
However, in the $Q=e$ regime  ($V_g\gtrsim-0.3425\;\mathrm{V}$),
we still measure a large noise which cannot be interpreted by
standard shot noise which is suppressed by the $(1-P)$ factor
\cite{Kumar96}. $S(2 \pi f_d)$ can even approach its
maximum value in this domain. In addition, $S(2 \pi f_d)$ exhibits
oscillations as function of $V_g$  which minima coincide with the
center of the diamonds of Fig.1.c. and maxima with their edges.

To understand the experimental results, we rely on a toy model of
the electron source. The period $T$ of
the excitation signal is divided in units of $\tau_0$, the time
needed for electrons to make one round trip inside the dot. When
promoted above the chemical potential during the first half-period
of the drive, the electron attempts to escape with probability $b$
every $\tau_0$. If it escapes, no additional electron is allowed
to escape, and a hole can be emitted during the next half-period
following the same rules. If the electron does not escape, the
emission of the hole is forbidden. The average current
computed in this model reproduces the exponential decay on a time
$\tau= \tau_0 \times (1/b-1/2)$ with an averaged emitted charge
per half period $Q=P.e$, $P=\tanh\big(T/4\tau(b)\big)$. The two contributions $C_1(t')= \overline{\langle
I(t)I(t+t')\rangle}^t $ and $C_2(t')= \overline{\langle
I(t)\rangle \langle I(t+t')\rangle }^t$ to the current
fluctuations $C=C_1-C_2$ calculated using the toy model have been plotted on Fig.3.a. in the case
of unit emission probability $P \simeq 1$ ($b=0.2$). $C_2$ is the
product of the statistical averages of the current, it reproduces
alternating peaks centered on electron and hole triggers ($t'=n
\times T/2$) and of typical width given by the average escape time
$\tau$. For times $t'\geq T/2$, $C_1$ equals $C_2$ reflecting the
absence of correlations between between two
successive electron/hole emissions. However, on short times $t'< T/2$, $C_1$
differs strongly from $C_2$: $C_1(t') \propto \delta(t')$ is a
Dirac peak proportional to $P$. As stated before, this is the
hallmark of a single particle emitter: the emission of an electron
cannot be followed by that of another one. This result can be
extended beyond this toy model. In full generality, considering
the emission of a single particle of charge $e$, one can show that
$\langle I(t)I(t+t')\rangle = e\langle I(t)\rangle \delta(t')  $.
If this perfect emission is triggered with period $T$, we have,
after averaging the time $t$ on one drive period,
$C_{1}(t')=\frac{e^2}{T} \;\delta(t')$. In our case, as one
electron and one hole are emitted at each period $T$, we get for
the perfect emitter $C_{1}(t')=\frac{2e^2}{T} \;\delta(t')$ with Fourier transform  $S_1(\omega)= 4e^2f_d$.  $C_2(t')$,
can be computed assuming only the exponential relaxation of the average current
$\langle I(t) \rangle$, we then get  $C_2(t') = \frac{e^2f_d}{\tau}
e^{-|t'|/\tau}$ with Fourier transform given by
$S_2(\omega)=\frac{4e^2 f_d }{1+(\omega \tau)^2}$. Their
difference $S(\omega)$ then reads
\begin{equation}
S_{jitter}(\omega)=4e^2 f_d \times \frac{(\omega
\tau)^2}{1+(\omega \tau)^2}
\end{equation}
In this optimum regime, the current fluctuations are not caused by
the fluctuations in the number of particles emitted between two
triggers, but are entirely determined by the quantum uncertainty
on the emission time of a single charge. This phase noise, which
we call quantum jitter, is the direct illustration that the
exponential decay of the average current, which looks like the
relaxation of a classical RC circuit, comes from the accumulation
of electrons emitted one by one with a random emission time coming
from the tunneling process. It can be used as a reference value
for perfect on-demand single particle emission and is fully
parameterized by the escape time $\tau$. Eq.(1) can thus be
experimentally checked by either varying $\omega$ at fixed $\tau$
or by varying $\tau$ at fixed $\omega$. Note that the quantum
jittering is encoded in the current correlations on times shorter
than the escape time $\tau$ or equivalently in the noise spectrum
at high frequencies $\omega \tau \approx 1$. To reach
subnanosecond time scales relevant for phase coherent electronics,
one needs the use of GigaHertz frequencies, in our case, we have
$\omega=2\pi f_d \approx 1/(100$ ps).

The short time behavior of $C_1$ and $C_2$ in the opposite limit
of long escape time, where the emission probability strongly
departs from one, $P<<1$ ($b=0.01$, $P=0.14$), has been plotted in
the inset of Fig.3.a. In that case, the contribution of the
average current $C_2$ is negligible, $C_1 \gg C_2 $ and the
current spectrum is white (except for very low frequencies) and proportional to $P$, $S_{shot}(2 \pi f_d)
=4e^2f_d\times P= e^2/\tau$. In this $P<<1$ limit, single charge
emission is a poissonian random process, the noise reflects these
random fluctuations in the emitted charge  and usual shot noise is
recovered.  Between the shot noise and quantum jitter limits,
$S(\omega)$ is also fully parameterized by the escape time $\tau$
and can be numerically evaluated \cite{Thesemahe}. An analytic
derivation of the noise spectrum in all regimes was even provided
in a recent paper \cite{Albert}.

Fig.3.b. represents our current noise data $S(2 \pi f_d)$ as a function of $\tau$ for operating conditions close to the center of the current diamonds, and their comparison with our model. The agreement is excellent within the full range
of escape time with no adjustable parameter. In particular, the shot
noise limit $\tau \geq T/2$ and more importantly the
quantum jitter limit $\tau <<T/2$ reproduce \emph{quantitatively} our
experimental results. The observation of the quantum jitter limit
demonstrates on-demand emission of a single particle, without collateral excitations. This corresponds to the optimum operating conditions of the source.

As seen in Fig.2.b, the model described above accounts
quantitatively for all our experimental data except for operating
conditions at the edges of the diamonds in the short escape time
regime (maxima of the oscillations). At these points, the
experimental data systematically fall above the theoretical quantum
jitter limit, represented by the black curve. In these operating
conditions, the dot charge for the initial and final value of the
excitation is not quantized as an energy level is brought at
resonance with the Fermi energy. Multiple charge emission then
occurs causing an excess of the noise with respect to the quantum jitter limit. These working points can not be used for single particle emission.

To conclude, we have measured the high frequency current
autocorrelations of an on-demand single charge emitter. In
particular, we have observed a new type of intrinsically high
frequency noise related to the quantum uncertainty on the emission
time of single charges. When the noise reduces to this quantum jitter, a single particle is emitted with unit probability between two shots of the source.
The use of these correlation techniques on single electron beams can now be applied to more
 elaborate electron quantum optics experiments. For example, two electrons interferences have been predicted \cite{Sam2004} and observed \cite{Neder} using continuous streams of electrons generated by DC biased ohmic contacts. Here, two electron interferences between single charges emitted on demand could be probed in a Hong-Ou-Mandel \cite{HOM} type experiment where two electrons collide on a beam splitter. Perfect antibunching at the beam splitter outputs would reveal the indistinguishability of electrons emitted by two independent sources \cite{Ol'khovskaya,feve08b}.\\

{\noindent\small{\bf Acknowledgements } We thank Anne Denis for the
fabrication of the $120$ to $50$ Ohms impedance matching lines.\\

\noindent $^{\dag}$also at SPEC-CEA Saclay, France

\noindent $^{\ddag}$Electronic address :
feve@lpa.ens.fr

\end{document}